\documentclass[preprint,longauthor]{aastex701}

\makeatletter
\long\def\frontmatter@title@above{%
  \vspace*{-\headsep}%
  \vspace*{\headheight}%
  \vspace*{0.625in}%
}
\makeatother

\shorttitle{GRaTer-JAX}
\shortauthors{Kondapalli et al.}

\begin{document}

\title{GRaTer-JAX: An Accelerated Package for Debris Disk Modeling}

\author[0009-0004-9073-5103]{Mihir Kondapalli}
\affiliation{University of California, Santa Barbara, Santa Barbara, CA 93106, USA}
\email{mihirkondapalli@ucsb.edu}

\author[0000-0002-8984-4319]{Briley L. Lewis}
\affiliation{University of California, Santa Barbara, Santa Barbara, CA 93106, USA}
\email{brileylewis@g.ucla.edu}

\author[0000-0001-5082-7442]{Jaren N. Ashcraft}
\affiliation{University of California, Santa Barbara, Santa Barbara, CA 93106, USA}
\email{jarenashcraft@ucsb.edu}

\author[0000-0001-6205-9233]{Maxwell A. Millar-Blanchaer}
\affiliation{University of California, Santa Barbara, Santa Barbara, CA 93106, USA}
\email{maxmb.astro@gmail.com}

\begin{abstract}

\texttt{GRaTer-JAX} is a Python package for modeling debris disks, disks
of dust around stars that provide important clues about the formation,
structure, and evolution of planetary systems. \texttt{GRaTer}
(\textbf{G}enerator of \textbf{R}ing-like, \textbf{a}xisymmetric,
optically \textbf{T}hin dust disks for \textbf{r}egularized fitting) is
a debris disk modeling framework originally developed for generating and
fitting models of optically thin, axisymmetric dust disks. JAX is a
Python package for high-performance computing that implements
just-in-time compilation and auto-differentiation, enabling significant
speedups in forward modeling workflows.

\texttt{GRaTer-JAX} delivers a more powerful, efficient, and robust
debris disk modeling framework than previous tools. Its JAX-based
backend provides orders-of-magnitude speedups and analytic gradients,
while its intuitive and extensible API unifies forward modeling,
optimization, and inference within a single workflow. Combined with key
additions to its disk model, it is capable of more advanced debris disk
modeling, enabling research that was previously computationally impractical such as
explorations of additional morphological parameters and disk composition
via scattering phase functions.

\end{abstract}

\keywords{Python; astronomy; debris disks; JAX; machine learning}

\section{Introduction}\label{statement-of-need}

\begin{figure}[htb!]
\epsscale{0.95}
\plotone{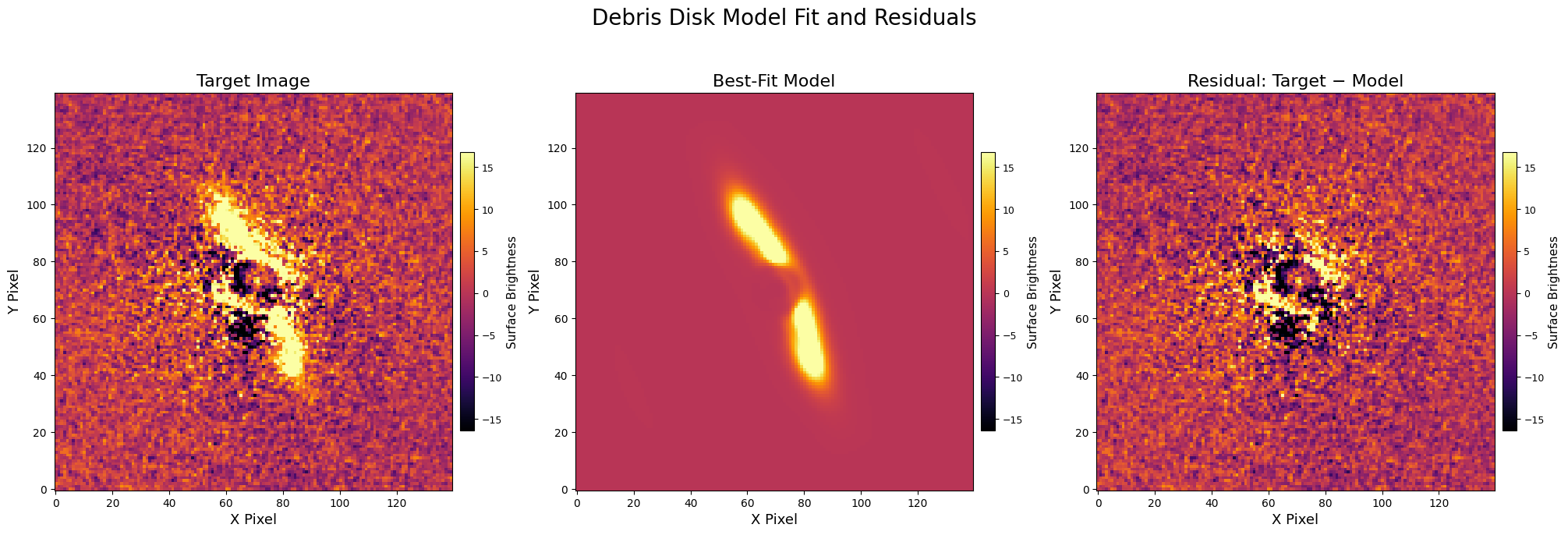}
\caption{GRaTer-JAX disk fit of HD 115600 in $H$-band polarimetry from
the Gemini Planet Imager debris disk survey \citep{Esposito2020},
illustrating the package's ability to reproduce disk morphologies from
real data.\label{fig:disk-fit}}
\end{figure}

Debris disks are circumstellar belts of dust and planetesimals, shaped
by a combination of stellar forces, dynamical interactions, and
collisional processes \citep{Hughes2018}. Their observed morphologies
provide key information for understanding the architecture, composition,
and evolutionary history of planetary systems. However, imaging
observations of such disks are often limited by noise, resolution, and
instrumental effects, making direct interpretation of data challenging
\citep{Hughes2018}. As a result, forward modeling (e.g., Figure~\ref{fig:disk-fit}) is
essential for extracting insight from observations.

For the past 25 years, the \emph{Generalized Radial Transporter
(GRaTer)} framework \citep{augereau1999} has provided an analytical
foundation for debris disk modeling and has been widely used to study
observed disk morphologies and infer physical disk properties (e.g.
\citep{Hughes2018}). A key component of the \texttt{GRaTer} model is the
scattering phase function (SPF), which describes how dust grains scatter
starlight toward the observer. Because the SPF is directly related to
dust grain properties, it can provide valuable insight into grain
composition, grain size distribution, and grain porosity. However, disk
modeling efforts have been historically limited by prohibitively long
computation times, requiring assumptions about disk properties to
simplify the models.

\texttt{GRaTer-JAX} was created to address the computational limits in
disk modeling by providing a fast, modern, open-source framework for
debris disk forward modeling and fitting. It is designed for
astronomers/researchers working on debris disk imaging who need faster
model evaluation, gradient-based optimization, and more flexible
parameterizations than are readily available in existing
implementations. By leveraging JAX \citep{Bradbury2018, Frostig2018},
including GPU acceleration, just-in-time compilation, and automatic
differentiation, \texttt{GRaTer-JAX} makes it practical to fit more than
25 parameters simultaneously, including flexible spline-based SPF
representations. This enables more detailed and expressive modeling and
broadens the range of debris disk structures and dust-scattering
behaviors that can be studied in practice.

\section{State of the Field}\label{state-of-the-field}

Historically, debris disk modeling has been substantially constrained by
computation time, restricting the number of disk parameters that can be
explored and reducing the flexibility of model fitting
\citep{Esposito2020}. Prior implementations of the SPF have also
generally relied on an empirical framework known as the
Henyey-Greenstein function \citep{henyey1941diffuse}, which is known to
be a poor descriptor of the observed scattering behavior of disks
\citep{Hughes2018}; this has limited previous models' ability to capture
more complex or unexpected scattering behavior present in real debris
disks. Existing implementations of the \texttt{GRaTer} framework, one of
the main examples being the Vortex Image Processing package
(\texttt{VIP}, \citep{GomezGonzalez2017}), have made debris disk forward
modeling more accessible, but they remain limited in several important
ways further described below. These limitations motivate the development
of a new implementation rather than a small extension of existing tools.

First, prior implementations are comparatively inefficient. As shown in
benchmark tests from the \texttt{GRaTer-JAX} repository, a basic
\texttt{VIP} debris disk model requires approximately 127 milliseconds
to generate on a 13th Gen Intel Core i9-13900K CPU. While this may
appear modest, it becomes burdensome in workflows that require thousands
of iterations, such as model fitting and parameter exploration.
\texttt{GRaTer-JAX} achieves a runtime of 2.36 milliseconds using an
NVIDIA RTX 4090 GPU, corresponding to a 54\(\times\) speedup for the
same model generation without sacrificing accuracy. As model complexity
increases, this speedup becomes even more pronounced. This increase in
speed also allows users to fit for additional parameters that were
previously inaccessible due to long computation times.

Second, existing implementations lack automatic differentiation. Without
access to analytic gradients, users must rely on numerical gradients,
which are both computationally expensive and less accurate. Benchmark
results show that computing a numerical gradient for a \texttt{VIP}
model with eight parameters requires approximately 2.04 seconds. This
severely limits gradient-based fitting workflows. In contrast,
\texttt{GRaTer-JAX} computes analytic gradients for a larger
twelve-parameter model in 16.8 milliseconds, corresponding to a 121
\(\times\) speedup. As with forward modeling, these performance gains
increase with model complexity.

Third, existing tools lack built-in support for modern optimization
workflows. \texttt{GRaTer-JAX} addresses this through its
\texttt{Optimizer} class, which provides built-in gradient descent and
Markov Chain Monte Carlo (MCMC) support. This framework also integrates
useful observational components such as throughput corrections for a
coronagraphic mask of the user's choice and convolution with a point
spread function (PSF). By bringing these capabilities into a unified
modeling workflow, \texttt{GRaTer-JAX} makes advanced fitting procedures
substantially more practical and accessible for debris disk studies.

All in all, \texttt{GRaTer-JAX} provides solutions to multiple existing
challenges in debris disk forward modeling, enabling faster, more
flexible, and more robust scientific analysis of debris disk systems.

\section{Software Design and Methodology}\label{software-design}

\begin{figure}[htb!]
\epsscale{0.95}
\plotone{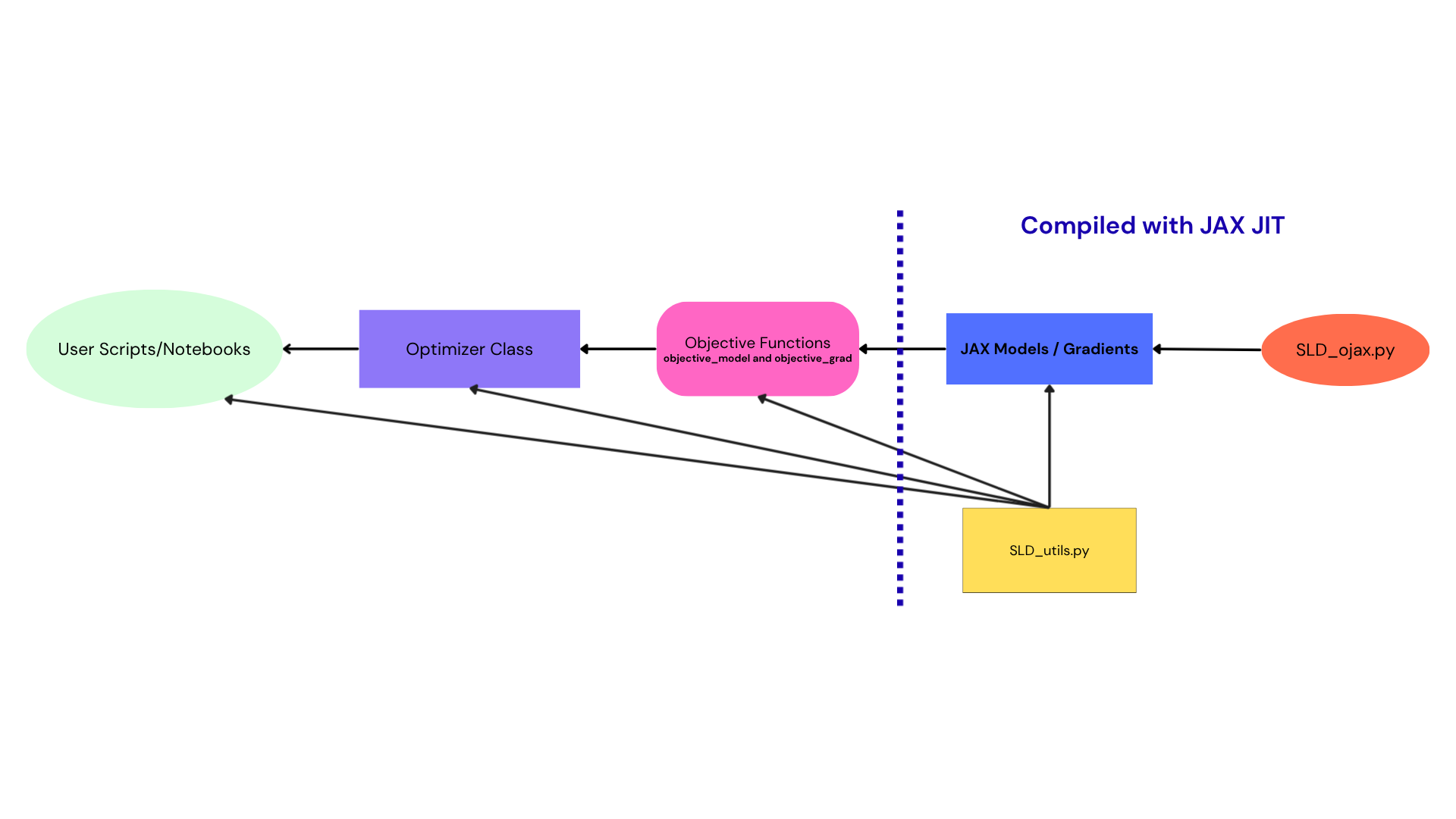}
\caption{High-level architecture block diagram, illustrating which
levels are and are not optimized via JAX just-in-time compilation.
\label{fig:architecture}}
\end{figure}

The software design of \texttt{GRaTer-JAX} was guided by a central
trade-off: exposing the full flexibility and performance of JAX while
providing an interface that is practical for astronomers using debris
disk models in real research workflows. A purely low-level JAX interface
would maximize flexibility, but it would also require users to manually
assemble model components, manage parameter transformations, and
understand JAX-specific implementation details. While powerful, that
approach would create a substantial usability barrier for researchers
whose primary goal is scientific modeling rather than software
engineering. On the other hand, a highly simplified interface could make
common workflows easier, but at the cost of limiting extensibility and
making it difficult to support more advanced models. \texttt{GRaTer-JAX}
was therefore designed as a layered system that balances performance,
usability, and extensibility; this architecture is visualized in a block
diagram in Figure~\ref{fig:architecture}.

\begin{figure}[htb!]
\epsscale{0.95}
\plotone{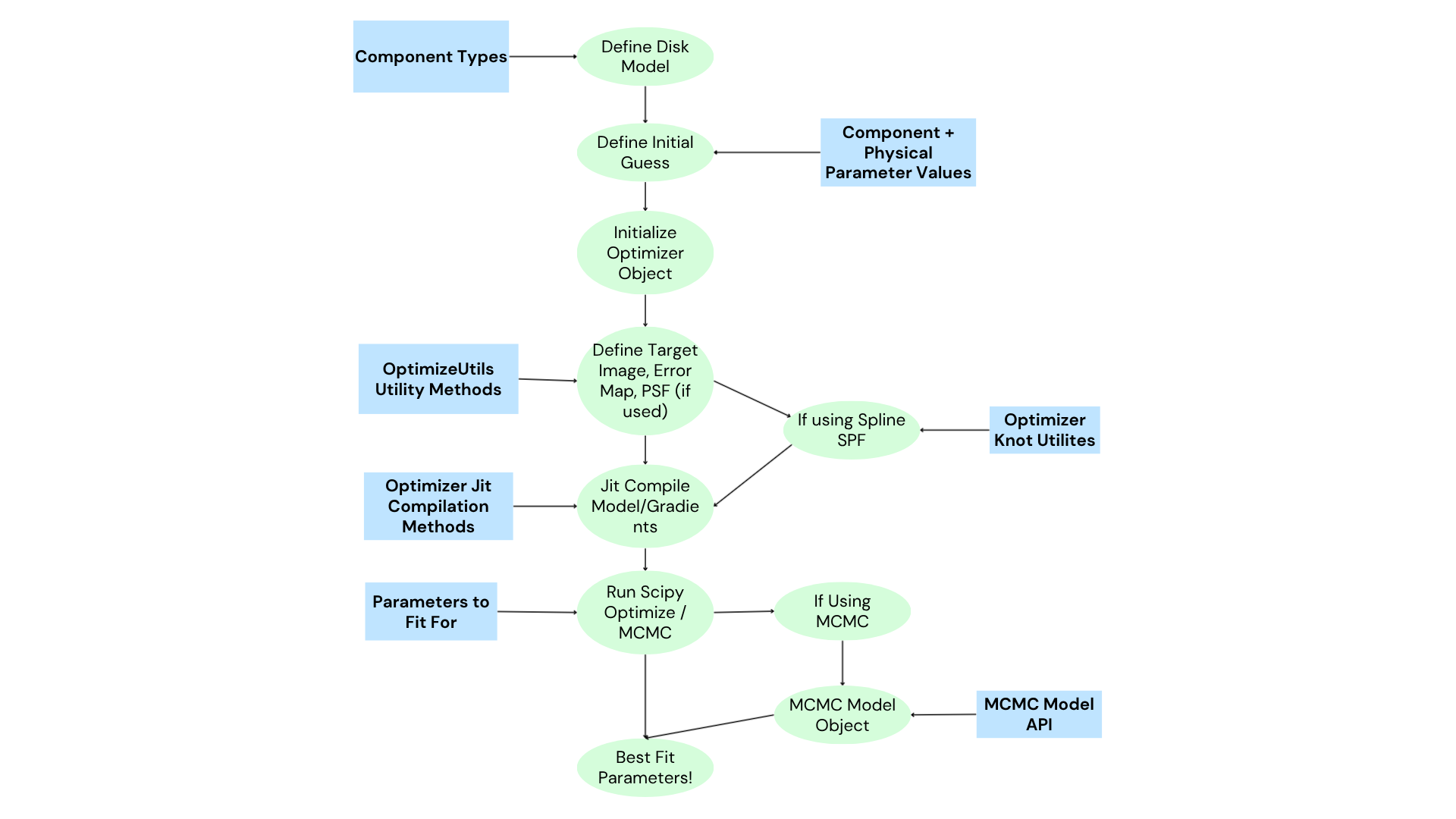}
\caption{Flow diagram for disk-fitting workflows, from initialization
through gradient descent and MCMC results.
\label{fig:fitting-workflow}}
\end{figure}

\texttt{GRaTer-JAX}'s layered architecture is key because debris disk
modeling is both computationally expensive and scientifically iterative.
Researchers often move from simple forward models to more advanced
fitting and inference. The low-level JAX layer provides speed and
differentiability, the objective-function layer improves usability, and
the optimizer layer connects the framework to real observational
analysis. Together, these choices make the software efficient, flexible,
and practical for research use. Figure~\ref{fig:fitting-workflow} summarizes the corresponding disk-fitting workflow, from model initialization through gradient-based optimization and posterior sampling.

\clearpage

\section{Research Impact Statement}\label{research-impact-statement}

\texttt{GRaTer-JAX}'s improved speedups and additions, as described
before, enable large scale precise parameter fitting. When models are
more powerful, researchers can avoid having to make fixed assumptions,
can analyze larger samples of disks in a uniform fashion to see
population-level trends, and can obtain more novel and trustworthy
results for individual disks. These, in turn, help us better understand
how debris disks evolve, what their compositions are like, and whether
unseen planets may be shaping them.

A key advancement enabled by \texttt{GRaTer-JAX} is the ability to fit
the SPF with a flexible spline, as opposed to the Henyey-Greenstein
function \citep{henyey1941diffuse}, which is parameterized by only the
asymmetry parameter \(g\) and has multiple known shortcomings
\citep{Hughes2018}. The free parameters of the SPF under our new
framework are the y-values of the spline knots; the x-values of the
spline knots are evenly spaced between \(0^\circ\) and \(180^\circ\), or
between \(\theta_{\rm sca,min}\) and \(\theta_{\rm sca,max}\) if using
inclination-based bounds, with a knot fixed to 1 at \(90^\circ\) to
normalize the SPF and avoid degeneracy with the flux scaling. The number
of knots \(n_k\) is determined by the user. With this new framework,
researchers have the potential to finally capture the true scattering
behavior of debris disks, therefore providing insight into their grain
properties.

This potential for more powerful models has already been realized in a
recent analysis of \emph{H}-band polarimetric debris disk images from
the Gemini Planet Imager (GPI)
\citep{macintosh2008gemini, LewisEtAl2026}, where \texttt{GRaTer-JAX} is
being used as the primary modeling engine. In \citep{LewisEtAl2026}, it
is used to forward-model the disk images, fit twelve or more debris disk
parameters together with spline-based scattering phase functions, and
sample posterior distributions with MCMC across a large target sample,
revealing population-level trends in planet formation/evolution.

Additionally, work on \texttt{GRaTer-JAX} has not only made this
modeling more accessible through its new, intuitive API and workflows,
but also through the development of a web app to help researchers
develop an intuition for how different disk parameters change the
observed morphology.
\href{https://scattered-light-disks.vercel.app/}{GRaTer Disk Image
Generator} provides researchers a simple interface to visualize debris
disk images using elements of the GRaTer-JAX package, allowing them to
workshop disks and try out different geometries before more detailed
fitting.

\section{Acknowledgments}\label{acknowledgments}

This material is based upon work supported by National Science
Foundation Astronomy \& Astrophysics Postdoctoral Fellowship Award
No.~2401654 for author BLL. Any opinions, findings, and conclusions or
recommendations expressed in this material are those of the authors
and do not necessarily reflect the views of the National Science
Foundation. Author J.N.A acknowledges support from NASA through Hubble
Fellowship grant HST-HF2-51547.001-A awarded by the Space Telescope
Science Institute, which is operated by the Association of Universities
for Research in Astronomy.

Packages used: \texttt{numpy}, \texttt{scipy}, \texttt{matplotlib},
\texttt{pandas}, \texttt{astropy}, \texttt{astroquery}, \texttt{poppy},
\texttt{stsynphot}, \texttt{synphot}, \texttt{pysiaf}, \texttt{emcee},
\texttt{arviz}, \texttt{corner}, \texttt{xarray}, \texttt{h5py},
\texttt{jax}, \texttt{jaxlib}, \texttt{jaxopt}, and \texttt{tqdm}

\clearpage
\bibliographystyle{aasjournalv7}
\bibliography{paper}

\end{document}